# Magnetodielectric coupling in a non-perovskite metal-organic framework

Tathamay Basu[1,*,#], Anton Jesche[2], Björn Bredenkötter[3], Maciej Grzywa[3], Dmytro Denysenko[3], Dirk Volkmer[3], Alois Loidl[1], Stephan Krohns[1]

[1]Experimental Physics V, Center for Electronic Correlations and Magnetism, University of Augsburg, D-86135 Augsburg, Germany

[2]Experimental Physics VI, Center for Electronic Correlations and Magnetism, University of Augsburg, D-86135 Augsburg, Germany

[3]Solid State and Materials Chemistry, Institute of Physics, University of Augsburg, D-86135 Augsburg, Germany

*e-mail: *tathamaybasu@gmail.com*

# Present address: Laboratoire CRISMAT, UMR 6508 du CNRS et de l'Ensicaen, 6 Bd Marechal Juin, 14050 Caen, France.

## Conceptual insights

Search for magnetodielectric and multiferroic materials, in which magnetism can be tuned by electric fields, is of considerable interest not only for basic research but also for functionalities for future technological applications. One recent and very promising route is to search for such properties in metal-organic-frameworks (MOFs). However, *until now, reports on multiferroicity in MOFs is restricted to a particular structure, so called perovskite structures, and also which mostly contain formate (HCOO) group* responsible for mediating magnetic exchange interaction. Using the enormous flexibility (of ligand and guest molecules) of MOFs system, *we have designed a new MOF which crystallizes in a **tetragonal structure** and consists of Co spin chains,* bridged by the functional ligand 4,4'-(2-nitrobenzol-1,4-diyl)bis(3,5-dimethyl-1H-pyrazol), *containing a **dipolar $NO_2$ group***.

Our result shows up that it might be possible to **introduce cross-coupling between magnetism and dielectric even by simply introducing a dipolar group in a designed MOF** in general, which opens up the possibility to design new MOFs in a wide variety of structure to achieve magnetodielectric coupling and multiferroicity. Therefore, our work yields a route to design new magnetodielectric MOFs which will help to achieve room temperature multiferroic for future potential applications in a long run.

### Abstract

**Multiferroicity and magnetodielectric coupling in metal-organic-frameworks (MOFs) is rare and so far restricted mainly to formate-based systems with perovskite structure. In the course of this work we designed a tetragonal framework [Co($C_{16}H_{15}N_5O_2$)], exhibiting spin-chains of $Co^{2+}$ ions, which are bridged by an organic linker containing a dipolar nitrobenzene moiety. This compound shows relaxor-like ferroelectricity at 100 K, which is followed by the onset of complex magnetic order at 15 K, indicative of weak ferromagnetism. The clear anomaly of the dielectric constant at the magnetic ordering transition indicates magnetodielectric coupling, which is also confirmed by magnetic-field dependent dielectric measurements. Weak ferromagnetism and magnetodielectric coupling, both probably result from a significant Dzyaloshinskii-Moriya interaction, which cants the spin structure and locally breaks inversion symmetry. We document that the introduction of dipolar nitrobenzene as building block in the crystal structure paths the way to design new multiferroic and magnetodielectric MOFs.**

## I. Introduction

In the past decade, multiferroic materials exhibiting coupled magnetic and electric order, in which magnetism can be controlled by electric fields and polarization by magnetic fields, have gained considerable interest because of potential application in storage devices and spintronics.[1,2,3,4] For an example, a single multiferroic (ferroelectric and antiferromagnetic) material with strong magnetoelectric (ME) coupling can be used to switch the magnetization via exchange bias at the heterostructure interface.[1] In such a device switching energy will be reduced by applying voltage instead of large currents.[1] However, single materials exhibiting multiferroicity, which allows switching via cross-coupling of magnetic and electric order are rare. Magnetism is related to spins and orbital momentum of partially filled $d$ or $f$ orbitals and proper ferroelectricity is a lattice phenomenon, which needs spatial inversion-symmetry breaking. The latter usually is favorable in compounds with empty $d$-shells ($d^0$, such as $Ti^{4+}$), because for these transition-metal ions (such as $Ti^{4+}$), off-centering can be induced by forming covalent bonds with neighboring atoms.[3,4] In particular, the most challenging tasks for such multiferroic materials are to increase the transition temperatures to ambient temperatures and to achieve a considerable coupling of magnetic and ferroelectric orders. Many efforts have been made to design multiferroic materials in different approach. Recently, multiferroicity has been established in metal-organic frameworks (MOFs), where transition-metal-ions with non-zero spin are connected by organic bridging ligands.[5,6,7,8,9] The enormous number of combinations of metal ions and organic linker molecules as well as their structural modification makes this material class promising to design MOFs for specific functions. Synthesis and characterization of new MOFs are driven by a variety of possible applications.[10,11,12,13,14] Their main advantages are flexibility in size and structure of the unit cell, and the possibility to implement a large variety of functional ligands and metal ions.[10,11] This allows designing MOF systems in order to find multiferroicity and magnetodielectric coupling (MDC). Another positive aspect of this class of materials is that MOFs are highly insulating, which is prerequisite to investigate multiferroic/MDC material. However, the observation of MDC and multiferroicity in MOFs is rare and so far was mainly restricted to a particular MOF systems, which crystallize in the perovskite structure and in most cases contain bridging formate (HCOO) groups, such as $[(CH_3)_2NH_2]M(HCOO)_3$ or $[NH_4]M(HCOO)_3$, where M corresponds to a $3d$ metal ion.[5,6,7,8,9] The functional linker group is responsible for mediating the super-exchange interaction between the magnetic ions.[7,8,9] The bridging ligand and its structure plays an important role for creating ferroelectricity.[7,15,16] In case of formate-based perovskite MOFs, it has been demonstrated that inversion symmetry is broken due to the distortion of the hydrogen bonds of the dimethylammonium cations ($[(CH_3)_2NH_2]^+$).[7,16] Systematic investigations of MOFs with different metal ions, suggest that the size of the ions influence the distortion of the lattice and thus strengthen ferroelectric order.[16,17] Till now, apart from perovskite-structure based MOFs, no investigations with the focus on multiferroicity were reported, though the huge flexibility in the design of MOFs may allow the synthesis of a suitable multiferroic material and will open up a vast possibility in this research field. Therefore, systematic studies on different coordination networks with various bridging ligands, metal ions, and structures are of high interest.

As a part of our research on functional MOFs, we have previously described a framework containing redox-active Co(II) centers, coordinated by linear 4,4'-(benzene-1,4-diyl)bis(3,5-dimethyl-1-pyrazolate) (bdpb$^{2-}$) linkers, named MFU-2 (Co(C$_{16}$H$_{16}$N$_4$), *M*etal-Organic *F*ramework *U*lm University-2).[18] MFU-2 crystallizes in tetragonal symmetry, with space group $P4_2/ncm$ (no. 138), unit cell parameters, $a = 18.576(4)$ Å, $c = 7.3728(18)$ Å, and volume, $V = 2544.0(10)$ Å$^3$. The structure is composed of cross-linked one-dimensional (1-D) cobalt(II) chains along the crystallographic *c*-axis (see figure 1a). In the present work, we have designed an isotypic variant of MFU-2, CFA-12 ([Co(C$_{16}$H$_{15}$N$_5$O$_2$)]), termed *C*oordination *F*ramework *A*ugsburg University-12), which contains dipolar nitrobenzene instead of non-polar benzene moieties as bridging linker (see figure 1). It crystallizes in the same space group as MFU-2, showing only slight variation of unit cell parameters, $a = 18.6070(8)$ Å, $c = 7.1635(5)$ Å, and volume, $V = 2480.2(2)$ Å$^3$. CFA-12 also contains 1-D Co(II) chains along the crystallographic *c*-direction (figure 1b and 1c). These Co(II)-ions are tetrahedrally coordinated by nitrogen atoms from the bridging 4,4'-(2-nitrobenzene-1,4-diyl)bis(3,5-dimethyl-1-pyrazolate) (NO$_2$-bdpb$^{2-}$) ligands, which have electron-withdrawing NO$_2$ groups, inducing a strong permanent dipole moment into the linker. Here, we report on synthesis and characterization of CFA-12, as well as on the magnetic and dielectric properties of this new MOF demonstrating multiferroicity and MDC.

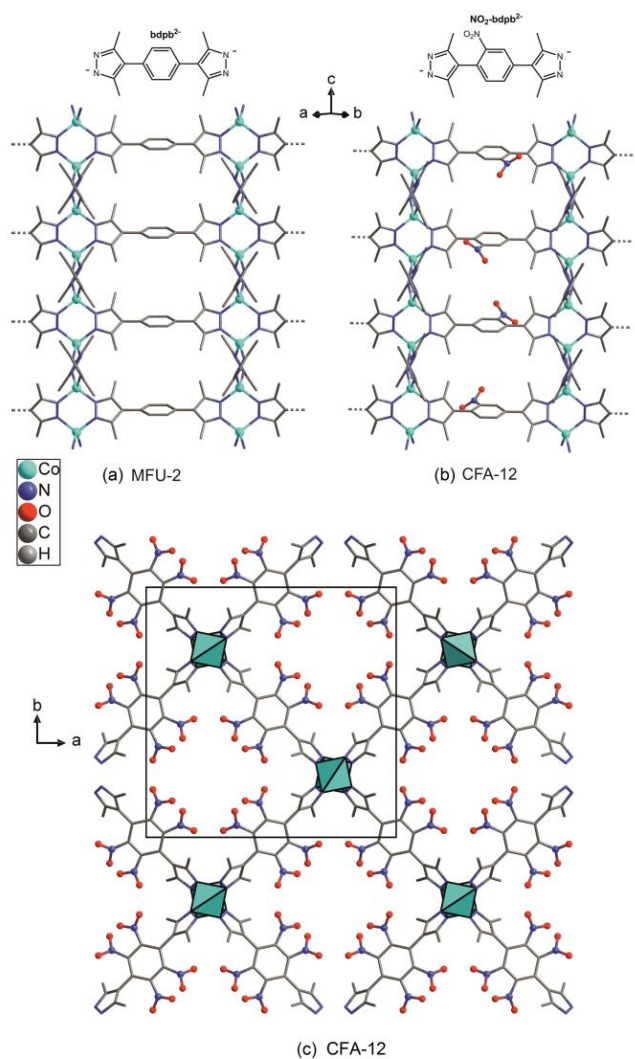

**Figure 1**: Schematic representation of selected parts of the crystal structures of (a) MFU-2 and (b) CFA-12 (along the *c*-direction). The hydrogen atoms were omitted for clarity (for details, see figure S1 in Supplementary Information). (c) Packing diagram of CFA-12 viewed along the *c*-direction. The site occupancy of the disordered $NO_2$ groups has a probability of 25 %. Disordered solvent molecules and hydrogen atoms were omitted for clarity.

## II. Experimental Methods

The 4,4'-(2-nitrobenzene-1,4-diyl)bis(3,5-dimethyl-1*H*-pyrazole) ($NO_2$-$H_2$-bdpb) ligand was synthesized by nitration of 4,4'-(benzene-1,4-diyl)bis(3,5-dimethyl-1*H*-pyrazole).[19] CFA-12 was prepared by a microwave-assisted solvothermal reaction starting from a Co(II) nitrate and NO$_2$-H$_2$-bdpb ligand in DMF, giving a violet microcrystalline product. Detailed mechanism of the synthesis as well as the analysis methods are given in Supplementary Information. The compound CFA-12 is well characterized by Fourier transform infrared (FTIR) spectra, mass spectrometer, thermogravimetric analysis (TGA) and X-ray diffraction analysis on both powder and single crystal specimen at various temperature. The temperature and magnetic field dependent magnetization measurements were performed using a Magnetic Property Measurement System (Quantum Design) equipped with a 7 T magnet. The dielectric properties for selected frequencies (ν) were analyzed using a LCR meter (E4980A, Agilent). An *ac* voltage of 1 V was applied on a powder sample, which was filled into a parallel-plate capacitor with an outer support ring made of Teflon.[20] Dielectric measurements were carried out in a Physical Properties Measurement System (Quantum Design) as a function of magnetic field and temperature. The measurements were performed during heating with a temperature rate of 0.5 K/min. In case of magnetic-field dependent measurement, the rate of the magnetic field is 50 Oe/s.

### III. Results

#### A. Structural analysis

In CFA-12, the asymmetric unit consists of one cobalt atom, two nitrogen, two oxygen, five carbon and three hydrogen atoms, which is ¼ of the NO$_2$-bdpb$^{2-}$ ligand. An Ortep-style plot of the asymmetric unit of CFA-12, with atom labels is shown in the figure S1 in Supplementary Information. CFA-12 features a non-interpenetrated microporous structure, constructed from chains of cobalt(II) ions bridged by NO$_2$-bdpb$^{2-}$ ligands and expanding along the *c*-direction. Each Co(II) center is coordinated by four nitrogen donor atoms, stemming from four different

NO$_2$-bdpb$^{2-}$ ligands. The local {CoN$_4$} coordination site adopts $D_{2d}$ point group symmetry, with all Co-N bond distances being equal (1.993(7) Å) and two out of six N-Co-N' angles being slightly smaller (105.6(4)°), compared to four larger angles (111.4(2)°). This values are in good agreement with those found in structurally related MFU-2 (Co-N: 1.992(5) Å, N-Co-N' angles: 106.6(3)° and 110.9(1)°). Adjacent cobalt(II) centers are bridged by NO$_2$-bdpb$^{2-}$ ligands to form 6-membered {Co$_2$N$_4$} rings, with a Co-Co distance of 3.6864(9) Å. This value is slightly larger than in MFU-2 (3.5818(3) Å). The phenyl rings of the organic ligands are twisted with respect to the plane, created by pyrazole rings of NO$_2$-bdpb$^{2-}$ ligands (47.9(2)°) and are parallel to each other in two neighboring ligands and almost parallel to the (001) plane. The positions of the NO$_2$ groups of the NO$_2$-bdpb$^{2-}$ ligands are statistically disordered. Owing to the high lattice symmetry, each NO$_2$ group appears with a site occupancy of 25 % at each of the four possible positions of the bridging benzene ring. The structure of CFA-12 creates rhombohedral channels running along the *c*-direction of the crystal lattice (see figure 1c). Taking the van der Waals radii of oxygen (1.52 Å) and hydrogen (1.2 Å) into account, the narrowest channel aperture between the hydrogen atom of the phenyl-ring and the oxygen atom of the NO$_2$-group is 6.31 Å, while the diameter is 8.27 Å. In the crystal structure of CFA-12, disordered Dimethylformamid (DMF) molecules occupy the channels. It was impossible to resolve and refine the positions of solvent molecules from the electron-density distribution. According to the crystallographic data, there is an electron count of 190 per unit cell, which corresponds to 4.75 DMF molecules in the unit cell of CFA-12.

Phase purity of CFA-12 was confirmed by X-ray powder diffraction (XRPD) at ambient conditions. Details of the synthesis and single-crystal structural refinement are described in the experimental method section in Supplementary Information. The experimental XRPD pattern of the sample (red line) is consistent with the simulated one (black line), as gleaned from the X-ray diffraction data (figure 2a). Differences in peak intensities are due to occluded solvent molecules. In addition, the thermal stability of CFA-12 was determined by variable temperature X-ray powder diffraction (VTXRPD) and thermogravimetric (TG) measurements. The TG profile (inset of figure 2b) of CFA-12 kept in nitrogen atmosphere exhibits a weight loss of 11 % between 300°C and 490°C, 28 % between 490°C and 650°C and 9.3 % between 650°C and 800°C. All three steps are connected with step-wise degradation of the compound. According to the VTXRPD data presented in figure 2b, the crystallinity of the sample decreases above 300°C, until the complete decomposition above 450°C is observed. The first weight loss step, where crystallinity is partially retained, might be associated with the loss of nitro groups (calculated value 12.5 %). The pore size distribution of CFA-12 is discussed in Supplementary Information.

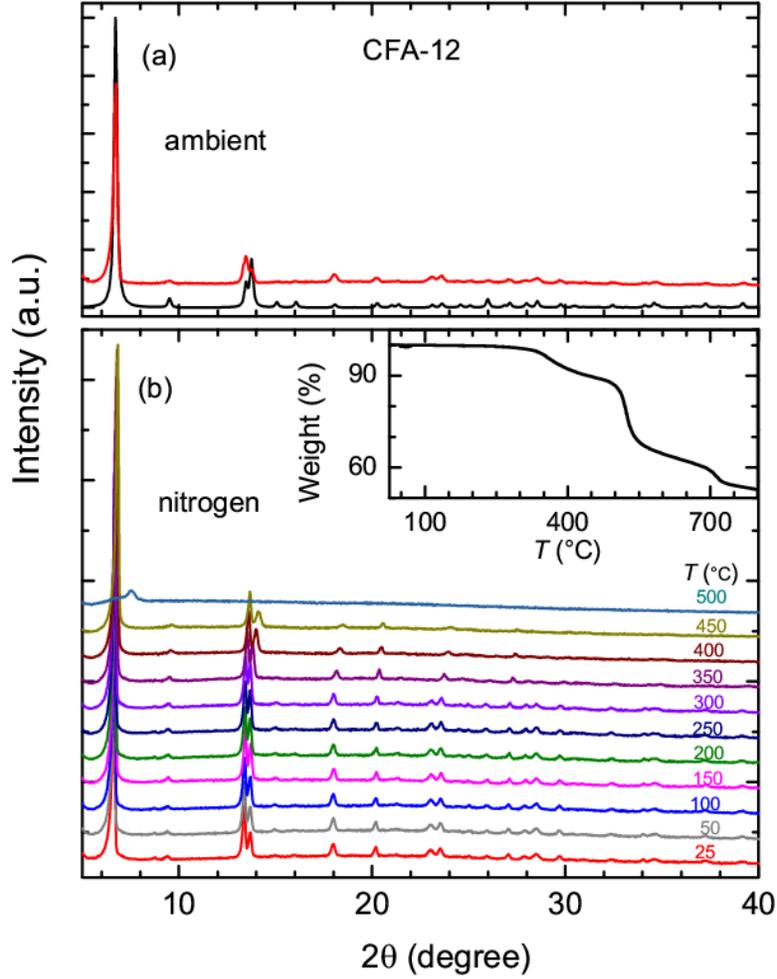

**Figure 2**: (a) Calculated (black) and measured (red) X-ray powder diffraction patterns for CFA-12. (b) VTXRPD plots of CFA-12 kept in nitrogen and sampled in a temperature range from 25-500°C. Inset shows the temperature-dependent weight loss of CFA-12 under flowing nitrogen gas.

### B. Magnetization

The *dc* magnetic susceptibility ($\chi = M/H$) in zero-field-cooled (ZFC) and field-cooled (FC) mode is shown in figure 3a for $\mu_0 H = 0.01$ T and 0.5 T. Below a characteristic temperature of $T_c \sim 15$ K, the increase in $\chi$ upon cooling gets significantly stronger than expected from Curie-Weiss behavior, indicating spontaneous magnetization (however with low magnetic moment), like a weak ferromagnet. For larger applied fields, this increase is less pronounced, becomes smeared out and shifts to higher temperatures, which is consistent with ferromagnetic (FM) ordering below this temperature. The comparison of *M/H* in ZFC and FC conditions reveals a bifurcation below 7 K (figure 3a). The ZFC curve exhibits a peak-like feature centered at $T_1 = 5.3$ K and an almost constant behavior for $T < 2.6$ K. The FC curve continuously rises with decreasing temperature. No bifurcation is observed in an applied field of $\mu_0 H = 0.5$ T. Figure 3b depicts the inverse susceptibility as a function of temperature ($T = 2 - 300$ K) for an applied magnetic field of $\mu_0 H = 0.5$ T. For $T > 100$ K a Curie-Weiss fit of the inverse susceptibility (figure 3b) yields an effective paramagnetic moment of $\mu_{eff} = 4.7$ $\mu_B$ per Co and a Curie-Weiss temperature of $\Theta_{CW} = -33$ K. As documented in figure 3b, the Curie-Weiss fit yields a perfect description of the inverse susceptibility, pointing towards well-defined local moments and antiferromagnetic exchange interactions. We also checked the Curie-Weiss fit considering a temperature-independent susceptibility ($\chi_0$) and found similar results with a small diamagnetic contribution, $\chi_0 \sim -10^{-9}$ m$^3$/mol (not shown). This is reasonable due to the diamagnetic contribution from the organic environment, usually observed in MOF systems. The negative sign of $\Theta_{CW}$ signifies dominant antiferromagnetic (AFM) exchange to be present in this compound. Also, $\Theta_{CW} > T_C$ indicates the possibility of short-range magnetic interactions above magnetic ordering. A closer look reveals that the magnetic susceptibility actually starts to deviate from Curie-Weiss behavior ($\chi^{-1} \sim T$) around 40 K, this is consistent with the presence of short-range magnetic interaction well above $T_C$. The experimentally observed effective moment is significantly larger than the spin-only value of Co$^{2+}$ (3.87 $\mu_B$ for $S = 3/2$) and indicates significant orbital contributions. Such large orbital contributions have also been observed in other cobalt based MOF compounds.[21,22]

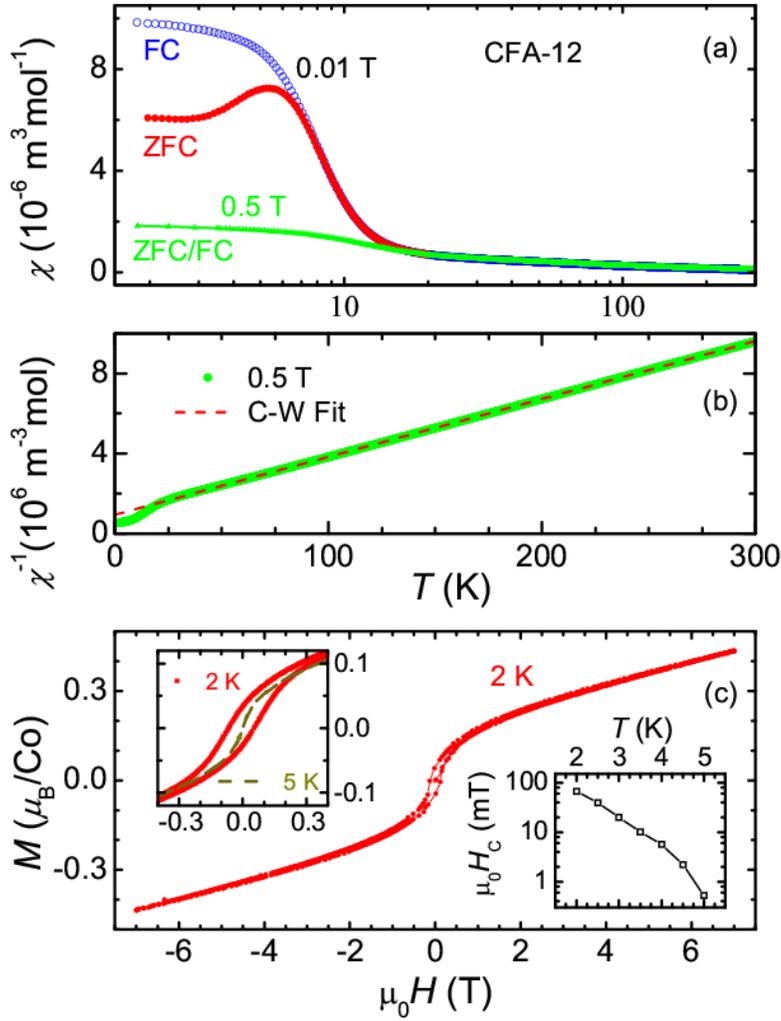

**Figure 3**: (a) *dc* susceptibility $\chi$ as a function of temperature in ZFC and FC conditions for external magnetic fields of 0.01 T and 0.5 T in the temperature range from 2 - 300 K on a semi-logarithmic scale. (b) Inverse susceptibility $\chi^{-1}$ measured at 0.5 T and Curie-Weiss fit in the paramagnetic region (100 - 300 K), which is extrapolated to low temperatures. (c) Isothermal magnetization is given as a function of magnetic field at 2 K to document the hysteresis loop. The left upper inset shows an enlarged view of the isothermal magnetization at 2 and 5 K, the right lower inset the temperature dependence of the coercive field for temperatures T < 5 K.

The isothermal magnetization *M(H)* is shown in figure 3c. The left inset of figure 3c depicts the behavior of *M(H)* in the vicinity of *H* = 0 T at *T* = 2 K and 5 K. A clear magnetic hysteresis shows up for 2 K, which supports the presence of FM order at low temperatures. No hysteresis effects are observed at 5 K. The coercive field as a function of temperature is plotted in the right inset of figure 3c. At $T = 2$ K the coercive field amounts to $\mu_0 H_C = 0.067$ T and decreases rapidly with increasing temperature. The magnetization at $T = 2$ K does not approach saturation even in the largest measured field of $\mu_0 H = 7$ T (figure 3c). The maximum value of $M$ ($T = 2$ K, $\mu_0 H = 7$ T) is roughly 0.43 $\mu_B$/Co, which is much lower than the magnetic moment of $Co^{2+}$ that is in the order of 3 $\mu_B$/Co assuming spin-only S = 3/2 and a g-factor close to 2.

How can these seemingly conflicting results, i.e. Curie-Weiss behavior with negative (antiferromagnetic) $\Theta_{CW}$ temperature and a large paramagnetic moment, but ferromagnetic spontaneous magnetization with small ordered moment, be reconciled? This certainly is possible assuming a canted antiferromagnet, where the weak FM moment arises from finite canting of mainly antiferromagnetically ordered spins. In addition, the strictly linear increase of the magnetization with increasing fields (figure 3c) also points towards a canted AFM. The increasing magnetization results from a decreasing canting angle on increasing external fields, until a collinear spin structure is achieved at high external magnetic fields. One might speculate that the cusp in the ZFC susceptibility and the bifurcation of FC and ZFC magnetization signal spin-glass behavior. However, the continuous increase of the FC curve below the bifurcation temperature is not characteristic for conventional spin-glass systems. In addition, we have not observed any appreciable frequency dependence in temperature dependent $ac$ susceptibility measurements in the frequency range of 1 Hz – 1 kHz (not shown). These facts exclude spin glass behavior. The splitting of FC and ZFC modes probably originates from domain wall effects in this ferromagnet when the applied magnetic fields are smaller than the coercive field.

We conclude that weak ferromagnetism (or canted AFM behavior) is realized in CFA-12. In many systems, weak ferromagnetism appears via a Dzyaloshinskii-Moriya (DM) interaction. DM interaction can only be active in systems without inversion symmetry. In this case one could speculate about the existence of polar order in the compound. Canted AFM structures are also suspected in some other MOF systems.[8,22,23] In these systems, anisotropic inter-chain and intra-chain interactions may play an important role in establishing this type of complex magnetism. Overall, the magnetic results certainly suggest the possibility of multiferroicity and of magnetodielectric coupling in this MOF compound under investigation.

### C. Dielectric and magnetodielectric properties

The temperature and magnetic field dependent dielectric properties were measured for frequencies from 1.15 to 100 kHz for temperatures T < 150K. Figures 4a and 4b show the temperature dependencies of real ($\varepsilon'$) and imaginary ($\varepsilon''$) parts of the complex dielectric constant in zero external magnetic field. The very low values of dielectric loss confirm the highly insulating nature of CFA-12 and also exclude an extrinsic Maxwell-Wagner-like[24] or magneto-resistive effect arising from leakage currents. In the investigated temperature and frequency regimes, we observe clear dielectric anomalies with significant dispersion effects: For 40 K < T < 110 K, the real part of the dielectric constant, $\varepsilon'$ reveals two distinct frequency dependent cusps (figure 4a), which are accompanied by peaks in $\varepsilon''$ (figure 4b). At 100 kHz the maximum of the dominant peak is close to 90 K and decreases to 84 K for 1.15 kHz. The frequency dependent behavior of $\varepsilon'$ is a typical signature of relaxor-like ferroelectric ordering.[9] However, an antiferroelectric compound could show similar features[25] and cannot be excluded.

To gain further insight into the relaxation dynamics of the dipolar degrees of freedom it is enlightening to analyze the frequency dependence of the loss maxima, which is shown in the inset of figure 4a. As function of frequency, the peak temperatures ($T_m$), obtained from the maxima in the dielectric loss, $\varepsilon''$, can best be fitted with an empirical Vogel–Fulcher-Tammann (VFT) law[26], $\nu = \nu_0 \exp[-E_a/(k_B(T_m-T_{VFT}))]$, shown in the inset of figure 4a in form of (ln$\nu$) vs. ($T_m$-$T_{VFT}$). Here $k_B$ is the Boltzmann constant and the parameters $\nu_0$, $E_a$ and $T_{VFT}$ are attempt frequency, activation energy and the characteristic VFT temperature, respectively. The latter ($T_{VFT}$) can be interpreted as static freezing temperature.[27] The parameters obtained from these fits are $\nu_0$ = 1.30 (±0.15)×10$^6$ Hz, $E_a$ = 17.3 (±0.2) meV and $T_{VFT}$ = 64 (±0.4) K. $\nu_0$ is much too low for a phonon-like frequency, but could correspond to a slow motion of the MOF network. The finite $T_{VFT}$ indicates static freezing of polar clusters. Below 64 K, CFA-12 behaves like a ferroelectric material with short-range polar order. Pyroelectric or polarization measurements are impeded due to the polycrystalline nature of the sample, which does not allow confirming the existence of ferroelectricity. A similar type of frequency-dependent dielectric relaxation at the onset of polar ordering is observed in the reported perovskite MOFs [(CH$_3$)$_2$NH$_2$]M(HCOO)$_3$, from which relaxor-type ferroelectricity has also been concluded.[9] Therefore, this dominant feature close to 90 K in CFA-12 most probably signals the transition into relaxor-like ferroelectricity at lower temperatures.

For 50 K < $T$ < 70 K a further frequency-dependent shoulder superimposes the low-frequency flank of the relaxor mode. This we identify as Debye-like relaxation,[28] which shifts to lower temperatures on decreasing frequencies, which is the typical behavior of dipolar relaxation, probably arising from partial reorientations of dipolar nitrobenzene groups carrying a dipolar moment. As function of frequency, the peak temperature ($T_P$) obtained from $\varepsilon''$, for this Debye-like relaxation, follows an Arrhenius law, $\nu = \nu_0 \exp[-E_a/(k_B T_P)]$, as shown in the inset of figure 4b. The fitting parameters are $\nu_0 = 7.9\ (\pm 1.2) \times 10^9$ Hz for the attempt frequency and $E_a = 95.6\ (\pm 1.1)$ meV, for the barrier.

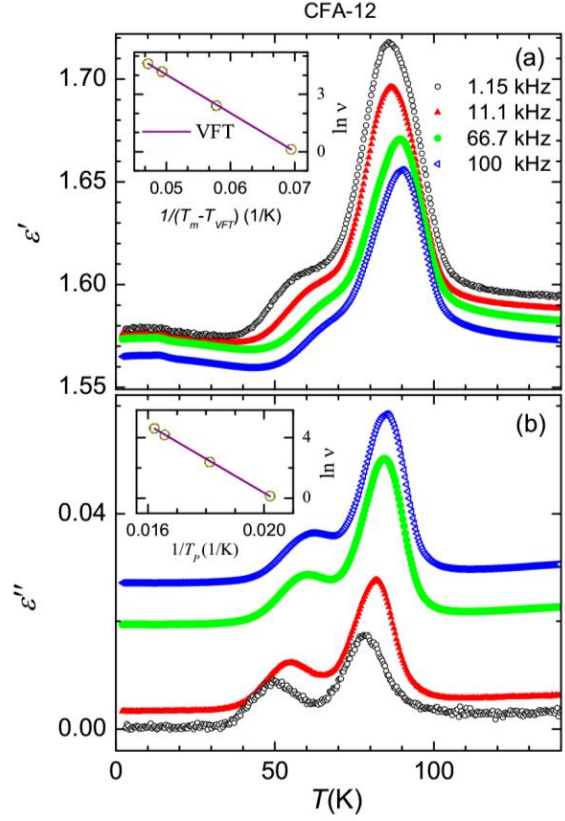

**Figure 4:** (a) Real and (b) imaginary part of dielectric constant of CFA-12 as a function of temperatures from 2 - 150 K for selected frequencies (1.15, 11.1, 66.7 and 100 kHz). Inset of (a) shows the frequency dependence of the peak temperature for the first relaxation process in 70 - 100 K in an Arrhenius representation introducing a freezing temperature $T_{VFT}$. Line is a fit using an empirical Vogel–Fulcher-Tammann law. The inset in (b) shows the Arrhenius behavior of the barrier of the secondary process.

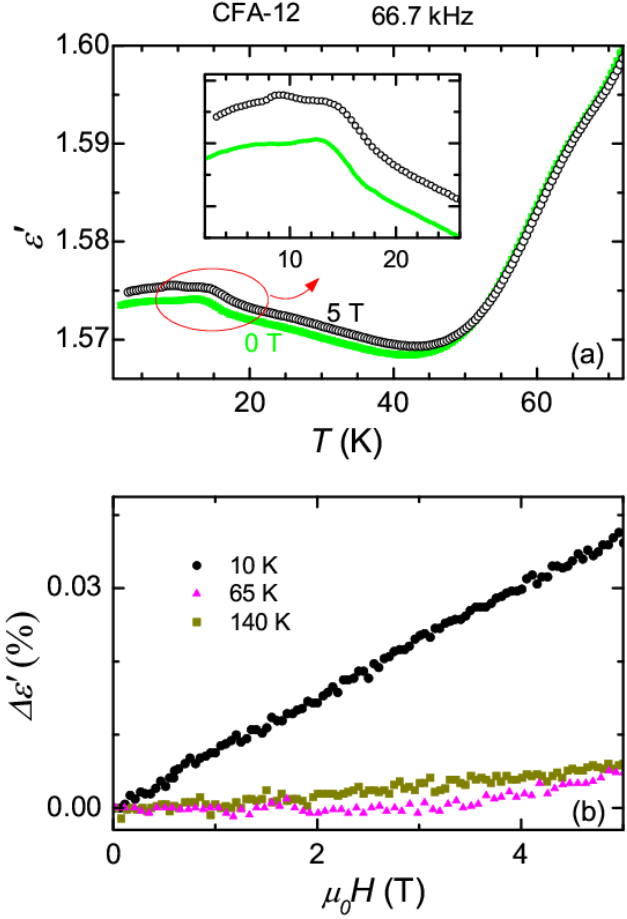

**Figure 5**: (a) Temperature dependence of the real part of dielectric constant of CFA-12 for a frequency of 66.7 kHz in zero magnetic field and in an external field of 5 T. (b) Magnetic-field dependent excess dielectric constant $\Delta\varepsilon'$ as measured at 66.7 kHz at temperatures of 10, 65 and 140 K.

Now we focus on the dielectric and magnetodielectric behavior at low temperatures. Figure 5a shows the temperature dependence of the real part of the dielectric constant $\varepsilon'(T)$ for temperatures < 70 K, as measured in zero external magnetic field, as well as in fields of 5 T, for frequency of 66.7 kHz. Below 40 K, the dielectric constant starts to increase approaching the magnetic phase transition. There is a clear hump-like anomaly close to 13 K, approximately coinciding with the onset of magnetic order. This dielectric anomaly, mimicking the magnetism, is also observed for other frequencies as well (figure S9 in Supplementary Information). At 5 T the temperature dependencies of dielectric constant above 40 K coincide as measured in zero field (figure 5a). However, a small but clear increase in $\varepsilon'$ appears below 40 K in the presence of the magnetic field. The MDC is even present above long-range magnetic ordering where (short-range) magnetic correlation starts to develop.[8,29,30] The change in dielectric constant ($\Delta\varepsilon'= [\{\varepsilon'(H)-\varepsilon'(0)\}/\varepsilon'(0)]$) as function of external magnetic field is documented in figure 5b. At 10 K, a strictly linear increase of the dielectric constant as function of magnetic field becomes apparent, while the changes in the dielectric constant are within experimental uncertainties for high temperatures. Therefore, figures 5 document finite linear MDC in CFA-12.

## IV. Discussion and Conclusions

The strength of MDC of CFA-12 (in powder form) is weak (~ 0.04% at 10 K), but comparable to that of reported single-crystalline perovskite MOFs (~ 0.05% at 10 K)[9] and many other bulk compounds including well-characterized frustrated spinel systems (~ 0.03%).[31,32] The coupling strength is usually higher in single crystals than in polycrystalline material. Within the metal-organic family, this system offers a good starting point for the synthesis of magnetodielectric materials outside the well-known perovskite structure. We cannot comment on the exact mechanism of the observed cross-coupling at present. The observed weak FM behavior is prone for the inverse DM interaction, therefore, we endorse such a mechanism to create lattice distortions without inversion symmetry that govern multiferroicity and magnetodielectric effects. The role of the inverse DM interaction on MDC (effect of the magnetic field on the ferroelectric order at the onset of magnetic ordering) has also been documented in organic-inorganic hybrid perovskite system $(NH_4)_2[FeCl_5(D_2O)]$ and $[CH_3NH_3][Co(HCOO)_3]$.[33,34,35] Although we favor

DM interaction as the driving mechanism for MDC. This cross-coupling may also be governed via magneto-elastic effects, as predicted in the MOF $[(CH_3)_2NH_2]Mn(HCOO)_3$.[8] Even, linear magnetodielectric effects in organic solids (in a spin-charge composite system) have been theoretically predicted in a recent report.[36] Such a mechanism is also possible in this Co-chain system. Further microscopic and theoretical investigations are highly warranted to shade further light on magnetism, magnetodielectric coupling and possible ferroelectric order in this new class of MOFs.[34,37]

To investigate the role of the polar nitrobenzene groups, we have performed magnetic and dielectric investigations on a MFU-2 compound, which has a similar structure without dipolar nitrobenzene groups. MFU-2 exhibits a similar magnetic ground state like CFA-12 (figure S10 and the discussion in Supplementary Information). However, MFU-2 does not show any kind of dielectric relaxation (figure S11 and the discussion in Supplementary Information). Therefore, the dipolar nitrobenzene groups play a dominant role in establishing dipolar moments and creating multiferroicity in CFA-12. We did not observe any clear feature in $\varepsilon'(T)$ at the onset of magnetic ordering in MFU-2. In addition no change of the magnetic field dependent dielectric constant within experimental resolution was observed. Figure 6 shows the change in dielectric constant as a function of magnetic field at 10 K for a fixed frequency of 66.7 kHz for the compound MFU-2 and CFA-12. This clearly document that CFA-12 exhibits appreciable MDC compare to MFU-2. Therefore, we conclude that MDC is negligible in the case of MFU-2 and inclusion of dipolar nitrobenzene in the same structure strongly influences the MDC. The disorder in the linkers is of statistic nature and the strong bonding of the linker does not permit re-orientations of the linkers at lower temperatures, which might lead to a structural ferroelectric-type phase transition. Certainly, dipolar $NO_2$ will have a strong influence on this mechanism.

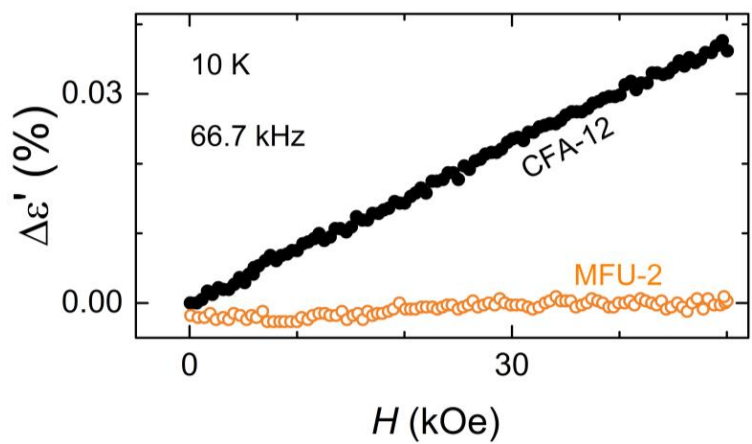

**Figure 6:** Excess dielectric constant Δε′ is given as a function of magnetic field for a fixed frequency of 66.7 kHz at 10 K for the compound MFU-2 (no $NO_2$ group) and CFA-12 (MFU-2 dipolar $NO_2$ group replacing H).

In summary, the structural, magnetic, dielectric and magnetodielectric properties of a new metal-organic-framework CFA-12 have been explored. CFA-12 has been designed by incorporating dipolar nitrobenzene-bdpb$^{2-}$ ligands into an earlier reported MFU-2 structure. This compound, containing spin-chains of Co(II) ions, bridged by nitrobenzene-bdpb$^{2-}$ ligands, exhibits magnetic ordering at ~ 15 K. The complex magnetic behavior is investigated in detail pointing towards the existence of weak ferromagnetic order. The dielectric properties indicate relaxor behavior around 100 K, suggesting a possible relaxor-ferroelectric phase, though, single-crystal work is needed to confirm ferroelectricity. Significant magnetodielectric coupling is established in this compound, with a coupling strength, which is comparable to perovskite-type carboxylate MOFs. Our investigation on MFU-2 and CFA-12 clearly reveals the significant role of dipolar nitrobenzene group on the dielectric and magnetodielectric behavior, whereas the magnetic properties are similar in these compounds. We demonstrate that cross-coupling between magnetic and electric order is not restricted to a particular perovskite MOFs and it is possible to introduce ferroelectricity and magnetodielectric coupling even by implementing a dipolar group in a designed structure. Therefore, this work opens up the possibility to design new magnetodielectric/multiferroic materials in a wide range of metal-organic-frameworks, apart from so-called perovskite MOFs, which may help on long terms to find a multiferroic and magnetoelectric material at ambient temperatures.

## Conflicts of interest

There are no conflicts to declare

**Acknowledgements**

Financial support by the German Research Foundation (DFG) via the Priority Program SPP 1928 "COORNETs", Grant No. JE 748/1 and the Transregional Collaboration Center TRR 80 are gratefully acknowledged. The authors thank Dr. C. Barkschat from the University of Augsburg for recording the mass spectrum. This work was supported by the BMBF via the project ENREKON 03EK3015.

# Supplementary Information

# Magnetodielectric coupling in a non-perovskite metal-organic-framework

Tathamay Basu[1]*, Anton Jesche[2], Björn Bredenkötter[3], Maciej Grzywa[3], Dmytro Denysenko[3], Dirk Volkmer[3], Alois Loidl[1], Stephan Krohns[1]

# 1. Properties of CFA-12:

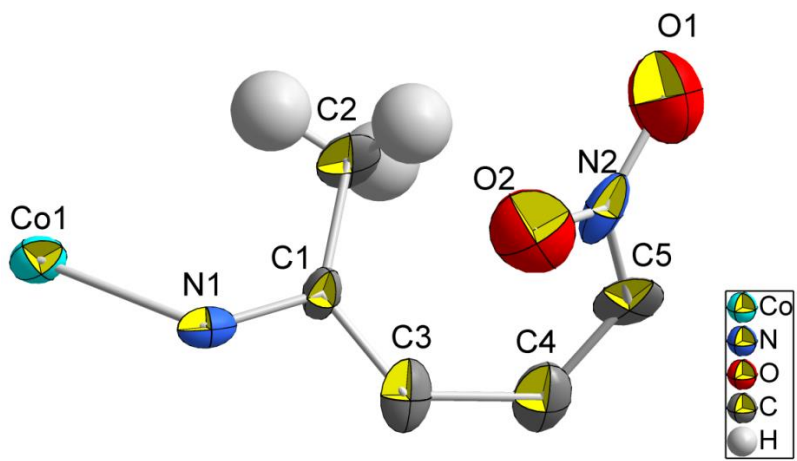

**Figure S1:** Ortep-style plot of the asymmetric unit of **CFA-12**. Thermal ellipsoids probability: 50 %.

## A. Sample synthesis

Commercially available reagents of analytical grade were used as received without further purification.

*Synthesis of 4,4'-(2-nitrobenzene-1,4-diyl)bis(3,5-dimethyl-1H-pyrazol) ($NO_2$-$H_2$-bdpb):*
4,4'-(benzene-1,4-diyl)bis(3,5-dimethyl-1*H*-pyrazole) (5.00 g, 18.8 mmol) was added in small portions at 0°C to 50 ml of concentrated sulfuric acid. Nitric acid (65%, 1.1 ml, 20.6 mmol) was added dropwise and the resulting solution was stirred for one hour at ambient temperature. Then the mixture was poured on ice (100 g) and neutralized by adding a solution of sodium hydroxide (75 g) in water (800 ml). The resulting precipitate was collected by filtration, washed with water and dried at 65°C to give the product (5.40 g, 17.4 mmol, 92 %) as a light yellow solid. M.p. 293-295 °C; IR: $\tilde{\nu}$ = 410w, 454w, 482w, 567m, 597w, 665w, 702w, 753s, 763s, 873m, 897w,

976w, 1008s, 1043s, 1112w, 1154m, 1245w, 1274m, 1308m, 1337s, 1359s, 1413s, 1437m, 1524vs, 1554s, 1585s, 2922m, 3022m, 3142m cm$^{-1}$; $^1$H-NMR (400 MHz, DMSO-d$_6$, 25°C): δ = 12.46 (s$_{br}$, 1 H, -NH), 12.40 (s$_{br}$, 1 H, -NH), 7.81 (d, 1 H, $^4J$ = 1.8 Hz), 7.62 (dd, 1 H, $^3J$ = 8.0 Hz, $^4J$ = 1.8 Hz), 7.41 (d, 1 H, $^3J$ = 8.0 Hz), 2.27 (s$_{br}$, 3 H), 2.24 (s$_{br}$, 3 H), 2.05 (s$_{br}$, 3 H), 1.96 (s$_{br}$, 3 H) ppm; $^1$H-NMR (400 MHz, DMSO-d$_6$, 80°C): δ = 12.21 (s$_{br}$, 2 H, -NH), 7.78 (s, 1 H), 7.60 (d, 1 H, $^3J$ = 8.0 Hz), 7.39 (d, 1 H, $^3J$ = 8.0 Hz), 2.26 (s$_{br}$, 6 H), 2.02 (s$_{br}$, 6 H) ppm; $^1$H-NMR (400 MHz, methanol-d$_4$, 25°C): δ = 7.82 (d, 1 H, $^4J$ = 1.8 Hz), 7.63 (dd, 1 H, $^3J$ = 8.0 Hz, $^4J$ = 1.8 Hz), 7.42 (d, 1 H, $^3J$ = 8.0 Hz), 2.32 (s$_{br}$, 6 H), 2.10 (s$_{br}$, 6 H) ppm; $^{13}$C-NMR (100 MHz, DMSO-d$_6$, 80°C) δ = 151.0 (q), 135.0 (t), 133.75 (q), 133.73 (q), 132.68 (q), 132.66 (q), 132.0 (q), 129.1 (t), 125.3 (t), 123.43 (q), 123.39 (q), 11.8 (p), 11.1 (p) ppm; $^{13}$C-NMR (100 MHz, CDCl$_3$/methanol-d$_4$ 9:1) δ = 150.2 (q), 142.5 (q), 142.0 (q), 134.5 (t), 133.5 (q), 133.4 (q), 132.7 (q), 132.6 (q), 129.1 (t), 125.4 (t), 124.2 (q), 124.1 (q), 11.29 (p), 11.23 (p), 10.77 (p), 10.74 (p) ppm; MS (ESI): m/z (%) = 312.15 (M + H)$^+$; elemental analysis: calc. (%) for C$_{16}$H$_{17}$N$_5$O$_2$ (311.3): C 61.72, H 8.63, N 22.49; found: C 61.59, H 8.78, N 22.39. $^1$H- and $^{13}$C-NMR are presented in figures S2-6.

*Synthesis of CFA-12 (microwave irradiation method):*

In a pyrex sample tube (35 mL), NO$_2$-H$_2$-bdpb (228 mg, 733 µmol) and Co(NO$_3$)$_2$·6H$_2$O (208 mg, 720 µmol) were dissolved at ambient temperature in DMF (4 mL). The tube was sealed and heated for 8 min at 120°C by a microwave synthesizer (CEM Discover S) at 300W. After cooling down to 20°C the product was filtered, washed with DMF (3x 20 mL) and CH$_2$Cl$_2$ (3x 10mL). Afterwards, the resulting solid was suspended three times in CH$_2$Cl$_2$ (3 mL) for 24 h (followed by centrifugation and decantation each time), and was subsequently dried in vacuum for 6 h at 25°C prior to further investigations to give phase pure CFA-12 (235 mg, 639 mmol, 87% based on NO$_2$-H$_2$-bdpb). Elemental analysis calc. (%) for Co(C$_{16}$H$_{15}$N$_5$O$_2$): C 52.2, H 4.1, N 19.0; found: C 51.7, H 4.4, N 19.2. The IR spectrum of CFA-12 is shown in figure S7.

*Synthesis of CFA-12 (solvothermal method):*

The crystals of CFA-12 suitable for single crystal X-ray analysis were synthesized by solvothermal synthesis: NO$_2$-H$_2$-bdpb (17.4 mg, 56 µmol) was dissolved in DMF (4 mL). To the solution Co(NO$_3$)$_2$·6H$_2$O (100 mg, 344 µmol) was added and dissolved. The resulting solution was placed in a heating tube (10 mL) which was closed with cap and heated at a constant rate of 0.10°C min$^{-1}$ to 120°C and then for 24 h at 120°C. The solution was subsequently cooled to 25°C and the violet crystals were taken for the single-crystal measurements.

## B. Physical Methods

Fourier transform infrared (FTIR) spectra were recorded with an attenuated total reflectance (ATR) unit in the range 4000–400 cm$^{-1}$ on a Bruker Equinox 55 FT-IR spectrometer. The following indicators are used to characterize absorption bands: very strong (vs), strong (s), medium (m), weak (w). Molecular mass was determined with a Q-Tof Ultima mass spectrometer (Micromass) equipped with an ESI source. Elemental analysis was performed on a Vario EL III,

Elementar-Analysensysteme GmbH. Thermogravimetric analysis (TGA) was performed with a TGA Q500 analyzer in the temperature range of 25–800 °C in flowing nitrogen at a heating rate of 10 K min$^{-1}$. Ar adsorption/desorption isotherms were measured with a Quantachrome I instrument. Adsorbed gas amounts are given in cm$^3$ g$^{-1}$ [STP], where STP = 100 kPa and 273.15 K. Prior to the measurement, the sample was heated at 180 °C for 5 h in high vacuum in order to remove occluded solvent molecules. Ambient temperature X-ray powder diffraction (XRPD) pattern was measured with a Seifert XRD 3003 TT diffractometer equipped with a Meteor 1D detector operated at 40 kV, 40 mA, CuK$_\alpha$ ($\lambda$ = 1.54247 Å) with a scan speed of 3 s per step and a step size of 0.02° in 2θ. The variable temperature XRPD (VTXRPD) data were collected in the 2θ range of 4-60° with 0.02° steps, with a Empyrean (PANalytical) diffractometer equipped with Bragg-Brentano$^{HD}$ mirror, PIXcel$^{3D}$ 2x2 detector and XRK 900 Reactor chamber (Anton Paar). The patterns were recorded in a temperature range from 25°C to 500°C. Temperature program between measurements: heating rate (0.5°C s$^{-1}$), then 10 min isothermal.

## C. Single-crystal X-ray diffraction

The crystals of CFA-12 ·1.2(C$_3$H$_7$NO) were taken from mother liquor and mounted on a MiTeGen MicroMounts. Several crystals were tested on the diffractometer. Unfortunately, most of the crystals scattered only up to ca. 32° 2θ (1.3 Å resolution). The best recorded data were obtained for a single crystal of CFA-12 ·1.2(C$_3$H$_7$NO) of approx. dimensions 35 × 20 × 18 μm$^3$. X-ray data for the single crystal structure determinations of CFA-12 ·1.2(C$_3$H$_7$NO) were collected on a Bruker D8 Venture diffractometer. Intensity measurements were performed using monochromated (doubly curved silicon crystal) MoK$_\alpha$ radiation (0.71073 Å) from a sealed microfocus tube. Generator settings were 50 kV, 1 mA. Data collection temperature was -173°C. APEX3 software was used for preliminary determination of the unit cell.[38] Determination of integrated intensities and unit cell refinement were performed using SAINT.[39] The structure was solved and refined using the Bruker SHELXTL Software Package.[40] Selected crystal data and details of structure refinement for CFA-12 ·1.2(C$_3$H$_7$NO) are provided in table S1. Complete crystallographic data for the structure reported in this paper have been deposited in the CIF format with the Cambridge Crystallographic Data Center, 12 Union Road, Cambridge CB21EZ, UK as supplementary publication no. CCDC 1513619. Copies of the data can be obtained free of charge on quoting the depository numbers. (FAX: +44-1223-336-033; E-Mail: deposit@ccdc.cam.ac.uk, http//www.ccdc.cam.ac.uk).

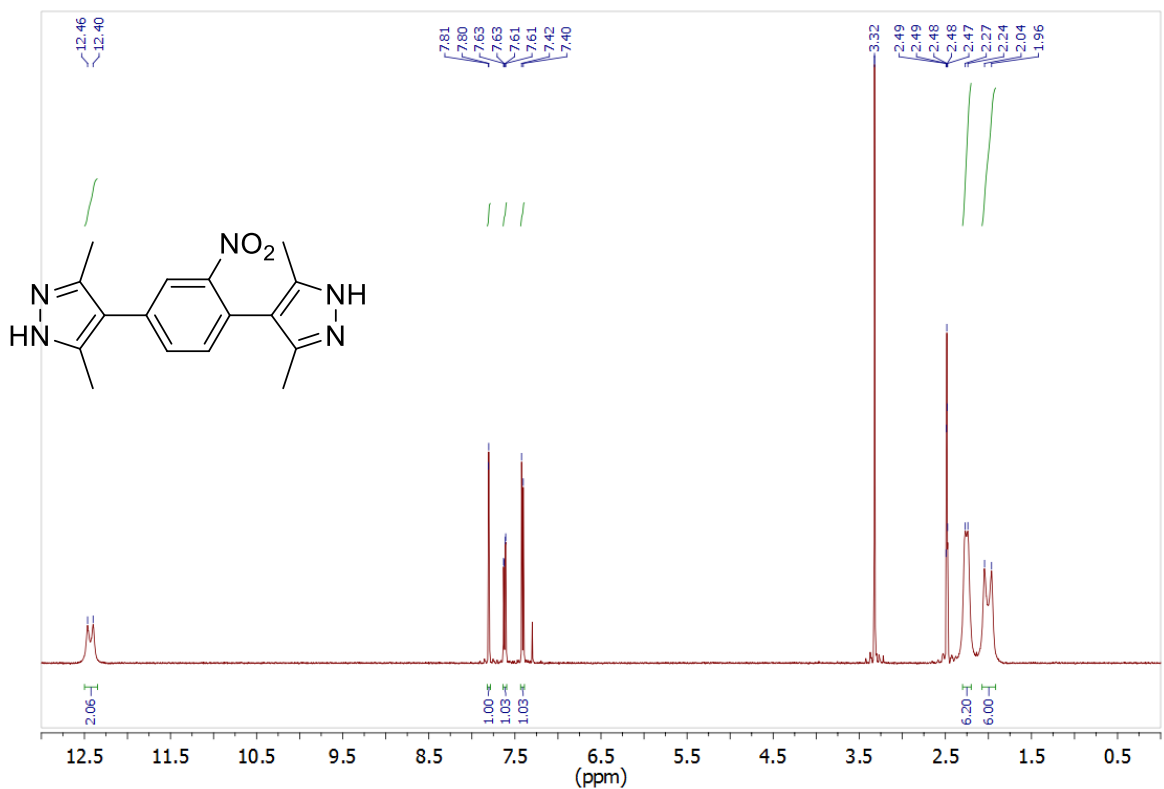

**Figure S2**: ¹H NMR spectrum (400 MHz, DMSO-d$_6$, 25°C) for 4,4'-(2-nitrobenzene-1,4-diyl)bis(3,5-dimethyl-1*H*-pyrazole) (NO$_2$-H$_2$-bdpb).

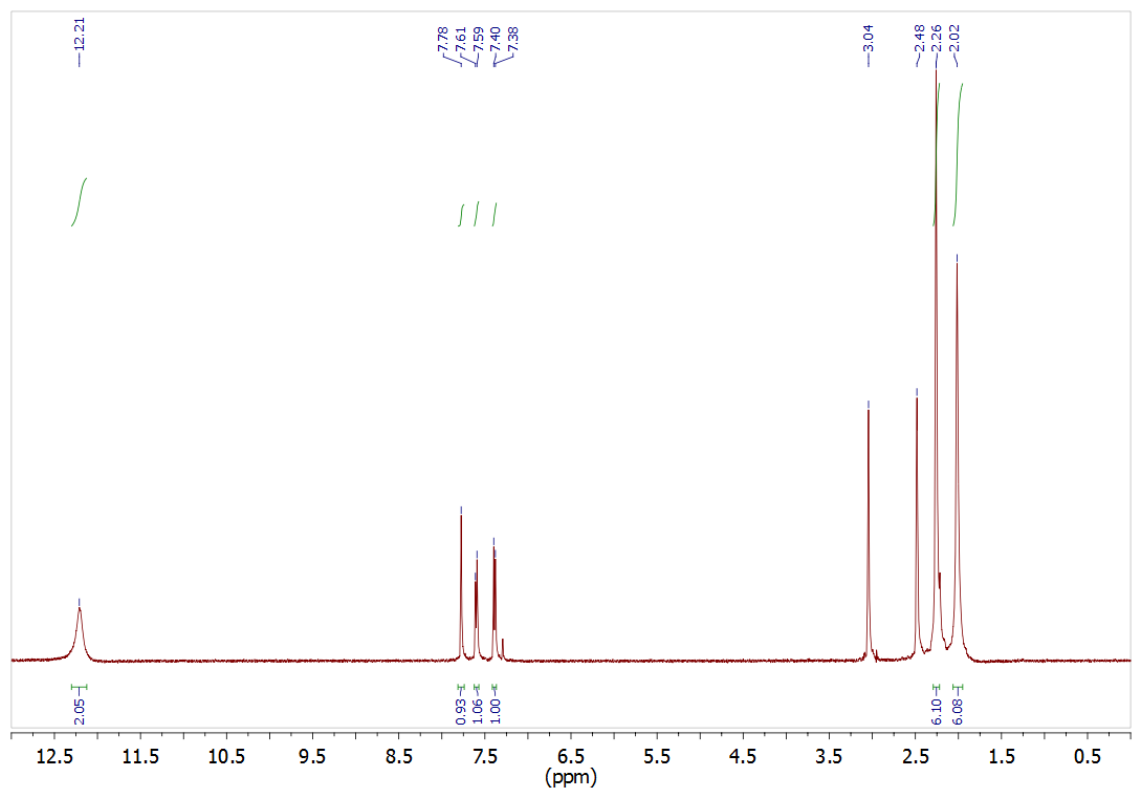

**Figure S3:** $^1$H NMR spectrum (400 MHz, DMSO-d$_6$, 80°C) for 4,4'-(2-nitrobenzene-1,4-diyl)bis(3,5-dimethyl-1*H*-pyrazole) (NO$_2$-H$_2$-bdpb).

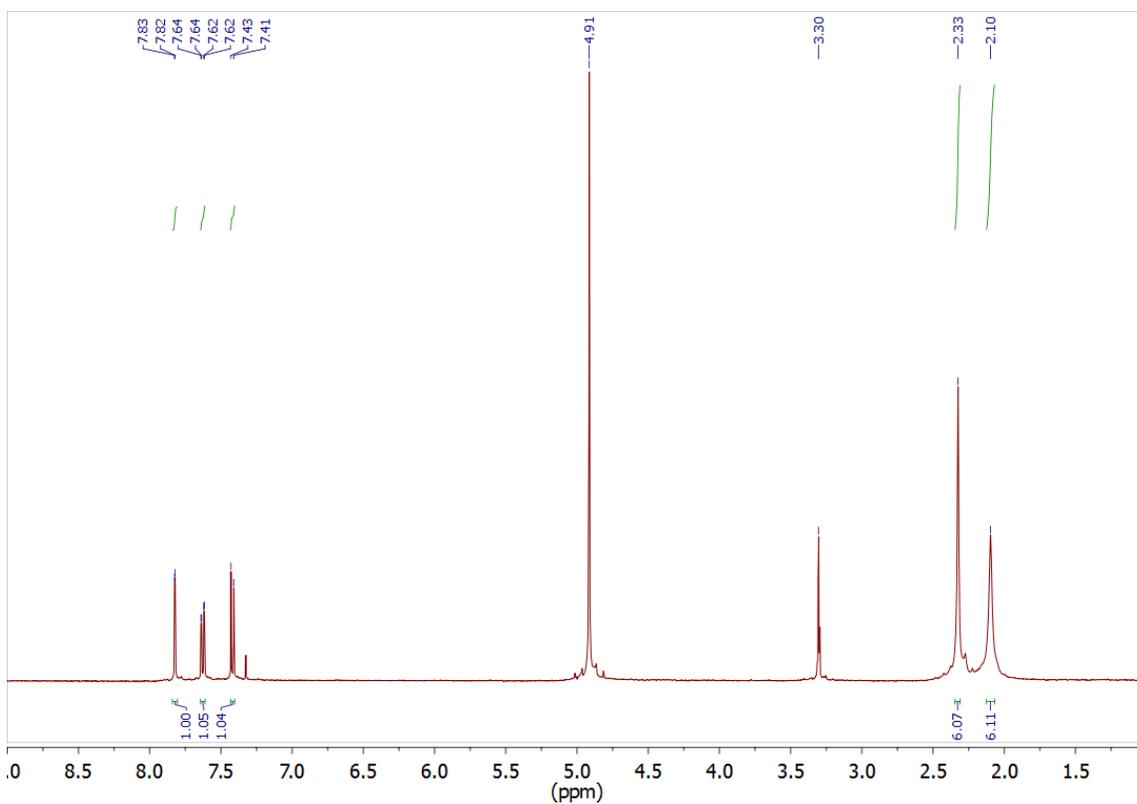

**Figure S4:** $^1$H NMR spectrum (400 MHz, methanol-d$_4$, 25°C) for 4,4'-(2-nitrobenzene-1,4-diyl)bis(3,5-dimethyl-1$H$-pyrazole) (NO$_2$-H$_2$-bdpb).

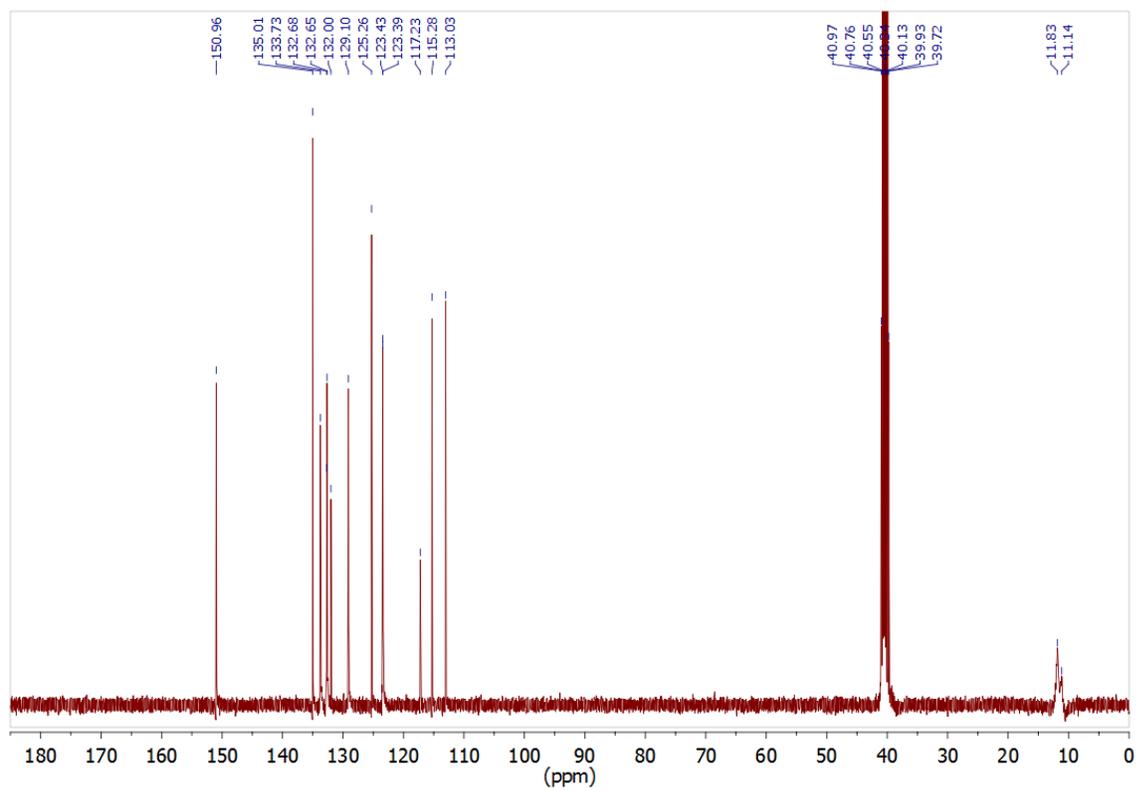

**Figure S5**: $^{13}$C-NMR spectrum (100 MHz, DMSO-d$_6$, 80°C) for 4,4'-(2-nitrobenzene-1,4-diyl)bis(3,5-dimethyl-1*H*-pyrazole) (NO$_2$-H$_2$-bdpb).

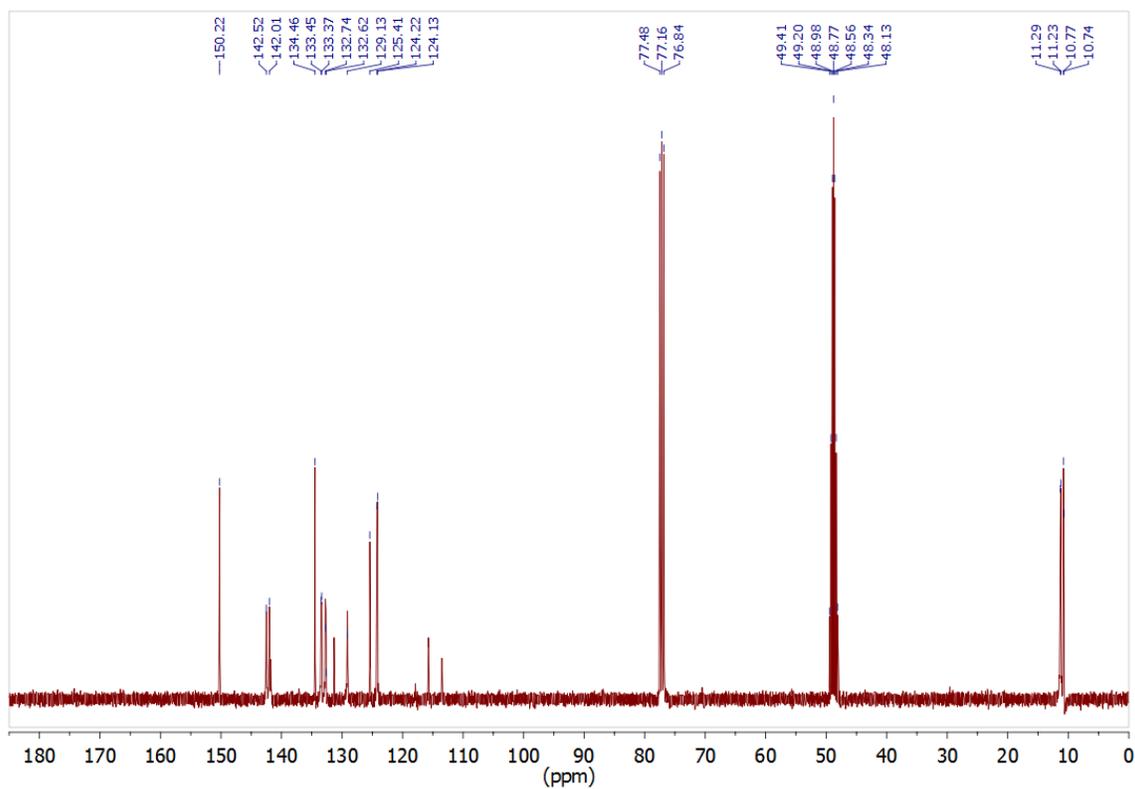

**Figure S6:** $^{13}$C-NMR spectrum (100 MHz, CDCl$_3$/methanol-d$_4$ 9:1, 25°C) for 4,4'-(2-nitrobenzene-1,4-diyl)bis(3,5-dimethyl-1*H*-pyrazole) (NO$_2$-H$_2$-bdpb).

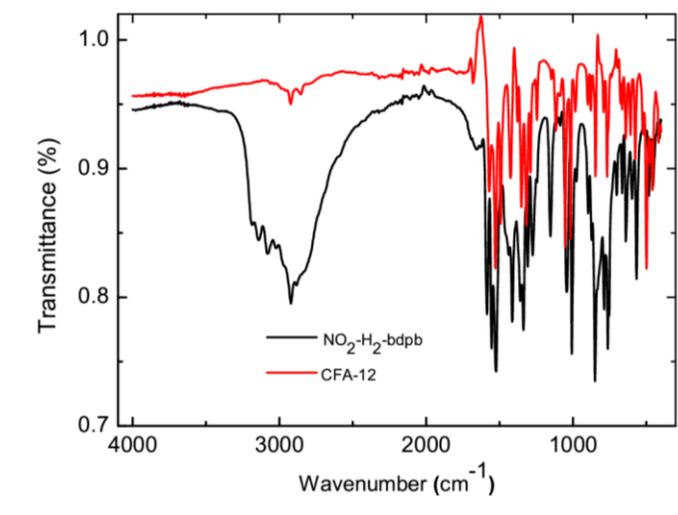

**Figure S7:** IR-spectra of 4,4'-(2-nitrobenzene-1,4-diyl)bis(3,5-dimethyl-1*H*-pyrazole) (NO$_2$-H$_2$-bdpb) and CFA-12.

**Table S1** Crystal data and structure refinement of CFA-12 ·1.2($C_3H_7NO$)

| Compound | **CFA-12** |
|---|---|
| Empirical formula | $C_{19.6}H_{23.4}CoN_{6.2}O_{3.2}$ |
| Formula | $C_{16}H_{15}N_5CoO_2·1.2(C_3H_7NO)$ |
| $M_r$/g mol$^{-1}$ | 455.98 |
| $T$/K | 100(2) |
| Wavelength/Å | 0.71073 |
| Crystal system | Tetragonal |
| Space group | $P4_2/ncm$ (no. 138) |
| $a$/Å | 18.576(4) |
| $c$/Å | 7.3728(18) |
| $V$/Å$^3$ | 2544.0(10) |
| $Z$ | 4 |
| $D_c$/g cm$^{-3}$ | 1.191 |
| $\mu$/mm$^{-1}$ | 0.704 |
| $F$(000) | 948 |
| θ Range/° | 3.10 to 26.03 |
| Refls. collected | 51515 |
| Refls. unique | 1280 |
| $R$(int) | 0.2147 |
| GooF | 1.206 |
| $R_1$ (I>2σ(I))$^a$ | 0.1652 |
| $wR_2$ (all data)$^b$ | 0.3729 |
| Largest diff. peak and hole/Å$^{-3}$ | 1.691 and -0.447 |

$^a R_1 = \Sigma||F_0|-|F_c||/\Sigma|F_0|$; $^b wR_2 = \Sigma[w(F_0^2-F_c^2)2]/\Sigma[w(F_0^2)2]^{1/2}$.

## D. Pore size distribution of CFA-12

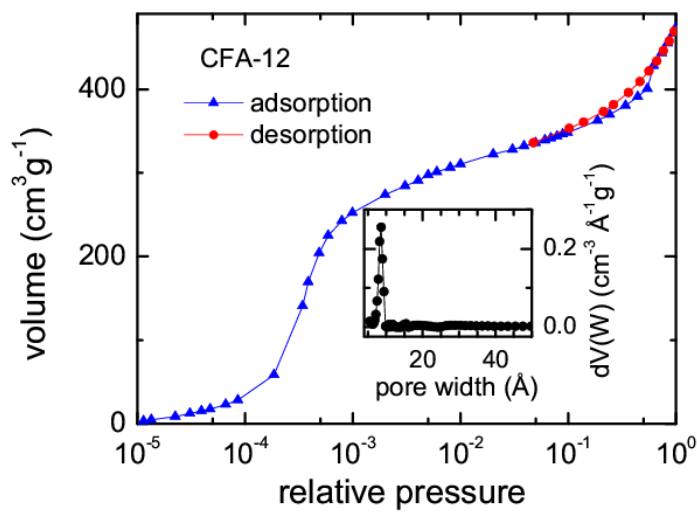

**Figure S8:** Ar adsorption/desorption isotherms for CFA-12 at 77 K; Inset: Pore size distribution for CFA-12.

The argon adsorption/desorption isotherm for CFA-12 at 77 K, presented in the inset of figure S8, is typical for microporous solids and reveals the Brunauer–Emmett–Teller (BET)[41] surface area of 1190 m$^2$ g$^{-1}$ and a total pore volume of 0.58 cm$^3$ g$^{-1}$. The slight hysteresis observed in the relative pressure range from 0.05-0.6 could be due to the flexibility of the framework (rotation of phenyl rings). The pore size distribution was calculated by using a non-local density functional theory (NLDFT),[42] implementing a carbon equilibrium transition kernel for argon adsorption at 77 K and based on a slit-pore model,[43] reveals a clear maximum at 8.6 Å (inset of figure S8), which is in good agreement with the crystallographic data (8.27 Å).

### E. Dielectric properties CFA-12

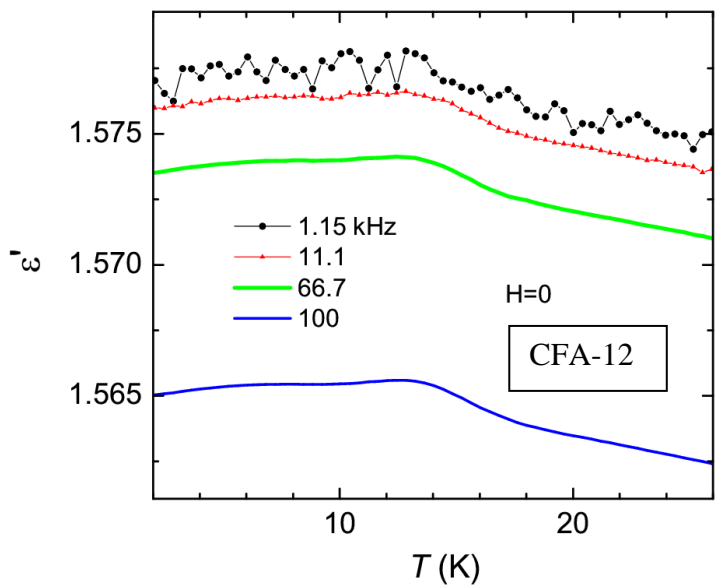

**Figure S9**: Magnified view of real part of dielectric constant as a function of temperature from 2-26 K for different frequencies between 1 and 100 kHz and for zero external magnetic field for the compound CFA-12.

## 2. Properties of MFU-2 (Co[C$_{16}$H$_{16}$N$_4$]):

### A. Magnetic properties of MFU-2:

The magnetic and dielectric investigation is performed on isostructural MFU-2. The dc susceptibility ($\chi=M/H$) as a function of temperature as measured at a field of 5 kOe is shown in figure S10. The compound shows similar magnetic behavior like CFA-12. The compound exhibits FM-like ordering below 20 K. The Curie-Weiss fit of the inverse susceptibility for T > 150 K yields an effective moment of $\mu_{eff}$ = 4.5 $\mu_B$ and a Curie-Weiss temperature of $\Theta_W$ = -37 K, which is nearly same as CFA-12, indicating AFM exchange interactions also in this compound. The ZFC-FC bifurcation is observed by application of low external magnetic fields (see left inset of figure S10), similar to bifurcation effects observed in CFA-12. We have also performed isothermal *M(H)* measurements at selected temperatures. The hysteresis at 2 K clearly confirms the FM nature of this compound. However, *M(H)* does not saturate at high fields, rather linearly varies with increasing magnetic field, which confirms the presence of AFM exchange also in this compound. The higher value of $\Theta_{CW}$ also manifests the short-range magnetic correlations arising from spin-chains of Co. Finally, we conclude, competing FM-AFM interaction is a common characteristic of this family, which is present in both systems, namely in MFU-2 and CFA-12.

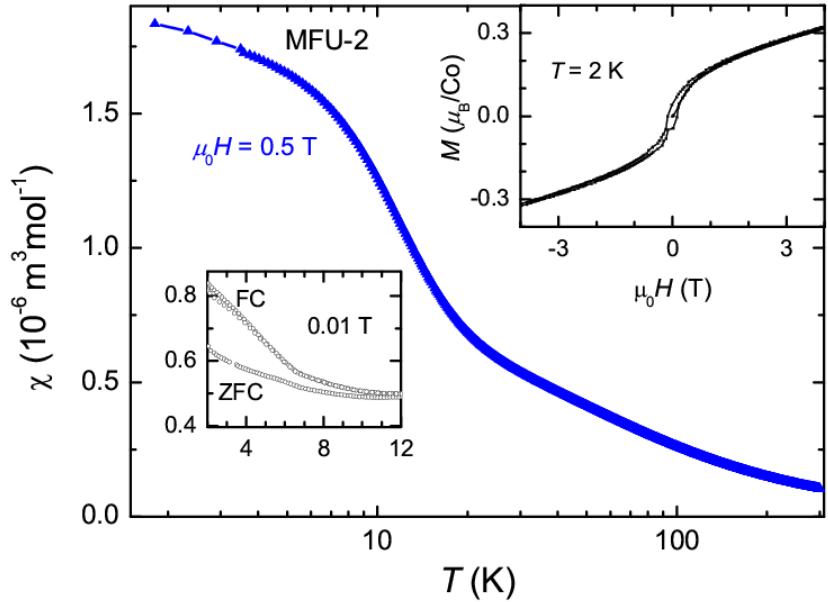

**Figure S10**: *dc* magnetization divided by magnetic field as a function of temperature for a magnetic field of 0.5 T at temperatures from 2 – 300 K; the left lower inset shows measurements at zero-field-cooled and field-cooled conditions for 0.01 T in the temperature range 2 – 13 K; the upper right inset documents the isothermal magnetization as a function of magnetic field at 2 K.

## B. Dielectric and magnetodielectric properties of MFU-2:

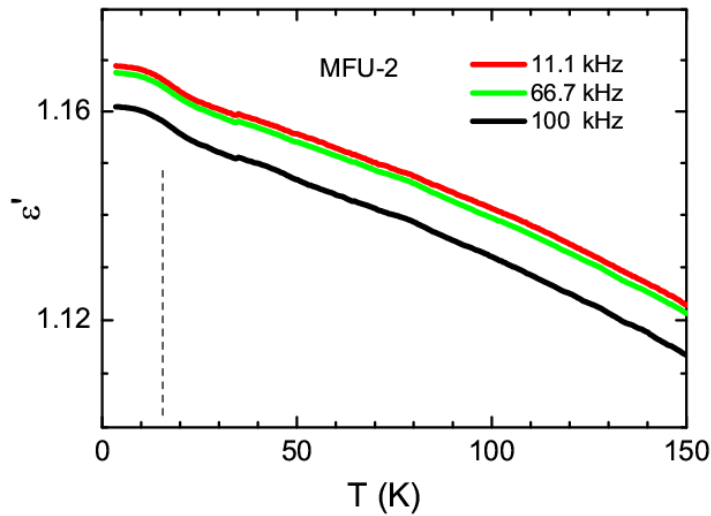

**Figure S11:** Real part of dielectric constant as a function of temperature from 2 – 150 K for some selective frequencies for the compound MFU-2. Dashed line indicates the magnetic transition.

The dielectric constant ε' as a function of temperature for the compound MFU-2 is shown in figure S11. No feature is observed at high temperatures. The slow and continuous increase of the dielectric constant with decreasing temperature could be due to volume changes by thermal expansion effects, which in turn decreases the thickness of the capacitor. Alternatively, it could be a genuine intrinsic temperature change of the dielectric constant. Though, there is no clear feature at the onset of magnetic ordering, a broadened feature is observed around 15 K. While this indicates the possibility of intrinsic magnetodielectric coupling in this compound, we did not observe any appreciable change in dielectric constant in the presence of a magnetic field, which suggest that the coupling strength is very weak in this compound (as shown in the Figure 6 in main article and discussed there). Therefore, the dielectric and magnetodielectric investigation on MFU-2 and CFA-12 clearly demonstrate the role of dipolar $NO_2$.